\documentclass{llncs}

\usepackage{stmaryrd}
\usepackage{bussproofs}
\usepackage{multirow}

\usepackage{llncsdoc}

\usepackage{mathptmx}

\usepackage{latexsym}
\usepackage{amsfonts}
\usepackage{amssymb}
\usepackage{amsmath}
\usepackage{graphicx}
\usepackage{color}

\usepackage{wrapfig,lipsum,booktabs}
\usepackage{semantic}
\usepackage{isabelle,isabellesym}
\usepackage{listings}
\usepackage{proof}

\usepackage{subcaption}
\usepackage{url}

\begin{document}

\title{A formalisation of the SPARC TSO memory model for multi-core machine code}  

\titlerunning{abbreviated title}  
%
\author{Zhe Hou\inst{2} \and David San\'{a}n\inst{1} \and Alwen Tiu\inst{3}  \and Yang Liu\inst{1} \and Jin Song Dong\inst{2}}
\authorrunning{Hou et al.} 
%
%
\institute{School of Computer Science and Engineering, Nanyang Technological University, Singapore
  \and
  Institute for Integrated and Intelligent Systems, Griffith University, Australia
  \and
  Research School of Computer Science, The Australian National University, Australia}

\maketitle

\begin{abstract}
  
  SPARC processors have many applications in mission-critical
  industries such as aviation and space engineering. Hence, it is
  important to provide formal frameworks that facilitate the
  verification of hardware and software that run on or interface with these processors. 
  This paper presents the first
  mechanised SPARC Total Store Ordering (TSO) memory model which
  operates on top of an abstract model of the SPARC Instruction Set
  Architecture (ISA) for multi-core processors. Both models are
  specified in the theorem prover Isabelle/HOL.
  We formalise two TSO memory models:
  one is an adaptation of the axiomatic SPARC TSO
  model~\cite{Sindhu1992,sparcv8}, the other is a novel operational TSO
  model which is suitable for verifying execution results. We prove
  that the operational model is sound and complete with respect to the
  axiomatic model. Finally, we give verification examples with
  two case studies drawn from the SPARCv9 manual.
  
  \end{abstract}

\section{Introduction}
\label{sec:intro}


As multi-core processors prevail in computers, it is important to
provide a formal specification of the instruction set architecture
(ISA) and weak memory model that establishes the precise principles
of concurrent low-level programs and the contract between hardware
and software. ISA provides the semantics of instructions and
processor operations, and it is essential in formal verification of the
correctness and security of micro-kernels~\cite{Klein2009,Gu2016}.
Weak memory behaviour is particularly important for low-level system
code such as synchronisation libraries, concurrent data structures,
concurrent program compilers, etc~\cite{Sewell2010}. The main purpose of such a specification is
to~\cite{Sindhu1992}

\begin{quote}
``Allow hardware designers and programmers to work independently, while still ensuring that any program will work as intended on any implementation.''  
\end{quote}

Sindu and Frailong also point out that a specification should be
formal so conformance to specification can be verified at some
level~\cite{Sindhu1992}. Interactive theorem proving allows one to
specify theories in rigorous mathematics and logic, and to reason
about the specification with machine assisted tools. Deductive
verification methods used in theorem provers enable the verification
of complex infinite-state systems, where automatic techniques such as
model checking struggle. As a result, formal verification projects,
such as the renowned seL4~\cite{Klein2009} and
CertiKOS~\cite{Gu2016}, rely on theorem provers and mechanised models
to provide a higher level of confidence that the formalisation is
correct. In our context, ``formal'' means that the model not only is
specified in mathematics, but also is mechanised in a theorem
prover.


The state-of-the-art on ISAs models cover different architectures such as Intel, AMD, SPARC, and PowerPC (references). Some of these formalizations also include weak-memory models to model multi-core architectures but as far as we are aware of, there are no formalisations of the weak memory model for the SPARC ISA. The multiprocessor SPARC architecture is adopted by the European Space Agency (ESA) to develop SPARC-based LEON multi-core processors in their space-crafts for critical missions~\cite{esaleon}. In order to formally verify concurrent software running in top of these CPUs down to the lowest layers of the execution stack, it is necessary to formalise the SPARC ISA and its weak memory model. To assist with the verification tasks, we need a model
that (1) supports SPARC ISA for multi-core processors, and (2) is formalised in a theorem prover. We focus on the SPARC TSO memory model since the critical software in our application uses TSO to avoid complex programs that require PSO. 
This work solves the above problems and serves as a
case study to the verification community for our specific needs.

We build upon the single-core
SPARCv8 ISA model of H\'ou et al.~\cite{hou2016}, which has been
tested against a LEON3 simulation FPGA board for correctness, and
develop a new SPARC ISA model for multi-core processors. The new ISA
model abstracts the detailed operational semantics in the SPARCv8 ISA
model into more general semantics 
while retaining the same operations in successful executions. Therefore, the previous experimental validation still holds for successful executions of the abstracted semantics. The new
semantics is more suitable to be used as an interface for memory
operations. The new ISA model is also an adaptation because various
considerations for multi-core processors are taken into account. We
drop the suffix ``v8'' for the abstract ISA model because we extend
the SPARCv8 model with features and instructions from the SPARCv9 architecture.
Specifically, we include the SPARCv9 atomic load-store instruction
Compare and Swap (CASA), which is not present in SPARCv8 manual
but is implemented on certain SPARCv8 processors. CASA is crucial for symmetric
multi-processors (SMP).

On top of the abstract ISA model, we give two TSO models: the first one is a
formalisation of the axiomatic SPARC TSO model~\cite{Sindhu1992,sparcv8}; the
second one is an operational TSO model which can be used to reason about
program executions. The integration of instruction semantics and weak memory
model is essential to support formal reasoning about concurrent programs, but
this problem is sometimes neglected in the weak memory
literature~\cite{Sewell2010}. We show that the operational TSO model is sound
and complete with respect to the axiomatic model. That is, every execution
given by the operational model conforms with the axioms, and every sequence of
memory operations that conforms with the axioms can be executed by the
operational model. Finally, we give two case studies based on the ``Indirection
Through Processors'' program and spin lock with CASA, both of which are drawn
from the SPARCv9 manual, to exemplify verifications on the order of memory
operations as well as on the result of execution. All the models and proofs in
this paper are formalised in
Isabelle/HOL\footnote{http://securify.sce.ntu.edu.sg/MicroVer/SparcTSO/TSO.zip}. 


\section{Related Work}
\label{sec:related}

An essential part of our work is the
formal model for SPARC instruction semantics. There has been much
work on formalising various instruction set architectures, but they
focus on instruction level modelling instead of memory operations. A
model of the SPARCv9 architecture is given by Santoro et
al.~\cite{Santoro1995}, but their model is not formalised in a
theorem prover. 
H\'ou et al.~\cite{hou2016} formalise the ISA for the integer unit of SPARCv8 single-core processors.
Their model can be exported for execution, and they have proven an instruction level
noninterference property for the SPARCv8 architecture. Fox et al.
give various models for ARM~\cite{Fox2003,Fox2010}, they also build a
framework for specifying and verifying ISAs~\cite{Fox2012,Fox2015}.
Goel et al. has a framework for building ISA models in
ACL2~\cite{goel2013}. There are also formalisations for compilers for
PowerPC, ARM, and IA32 processors~\cite{Leroy2006,compcertc}, and for
JVM~\cite{Liu2003,Atkey2005}. Our ISA model differs from the above
work in that we model multi-core processors.


There is an non-exhaustive list of
literature on relaxed memory models, but most of them do not consider
machine code semantics. Here we only discuss the most closely related
ones. Typically memory models appear in two forms: axiomatic model
and operational model. The axiomatic TSO memory model for SPARC is
given by Sindhu and Frailong~\cite{Sindhu1992}. This model is used in
the SPARCv8 manual~\cite{sparcv8}, and is later referred to as the
``golden memory model''~\cite{loewenstein2006}. Petri and
Boudol~\cite{petrithesis,BoudolP09} give a comprehensive study on
various weak memory models, including SPARC TSO, PSO, and RMO. They
show that the store buffer semantics of TSO and PSO corresponds to
their semantics of ``speculations''. Gray and Flur et
al.~\cite{Gray2015,Flur2016} have established axiomatic and
operational models for TSO, and their equivalence. Their work is also
integrated with detailed instruction semantics for x86, IBM Power,
ARM, MIPS, and RISC-V. They have developed a language called Sail for expressing
sequential ISA descriptions with relaxed memory models that later can be translated into Isabelle/HOL.
 However, the current set of modelled ISA does not include any variance for the SPARC ISA. Although it would have been possible to rewrite the semantics of~\cite{hou2016} in Sail, this language lacks some important features necessary for our work. First, Sail does not provide some low level system semantics such as exceptions and interrupts; second, their framework does not include an execution model for multi-core processors. 


Besides
Burckhardt's work, there are other tools and techniques developed for
verifying memory operations. Notably, Hangal et al.'s
TSOtool~\cite{Hangal2004} is a program for checking the behaviour of
the memory subsystem in a shared memory multiprocessor computer
aginst the TSO specification. Although verifying TSO compliance is an
NP-complete problem, the authors give a polynomial time incomplete
algorithm to efficiently check memory errors. Companies such as Intel
also actively work on tools for efficient memory consistency
verification~\cite{Roy2006}. Roy et al.'s tool is also polynomial
time and is deployed across multiple groups at Intel. A tool
specialised for SPARC instructions is developed by Park and
Dill~\cite{Park1995}.

There are also memory
models that are formalised in theorem provers, such as Yang et al.'s
axiomatic Itanium model Nemos in SAT solvers and
Prolog~\cite{Yang2004} and the Java Memory Model in
Isabelle/HOL~\cite{Aspinall2007}. Alglave et al.
formalised a class of axiomatic relaxed memory models in
Coq~\cite{Alglave2010}. Crary and Sullivan formalise a calculus in
Coq for relaxed memory models~\cite{Crary2015}. Their calculus is
more relaxed than existing architectures, and their work is intended
to serve as a programming language. A more related work is Owens, Sarkar,
Sewell, et al.'s formalisation of x86 ISA and memory
models~\cite{Sarkar2009,Owens2009,Sewell2010}. They formalise both
the ISA and relatex memory models such as x86-CC and x86-TSO in HOL
and show the correspondence between different styles of memory models.
It is possible to translate Gray and Flur et al.'s
work~\cite{Gray2015,Flur2016} to Isabelle/HOL or Coq code. However,
the resulting formal model would rely on the correctness of the
translation tool such as Lem~\cite{Mulligan2014}, which adds one more
layer of complication in our verification tasks.

\section{SPARC Abstract Instruction Set Architecture}
\label{sec:isa}

This section presents our abstract SPARC ISA model, which is an
abstraction and adaptation of the one of H\'ou et al.~\cite{hou2016}.
The previous model
is suitable for reasoning about
operations at instruction level, but it is too complex and detailed
to reason about memory operations. Hence we abstract their work into a
more general model with big-step semantics and less SPARC specific
features. Besides the non-memory-access instructions in the integer
unit, we focus on the following instructions for memory access: load
(LD), store (ST), swap memory and register content (SWAP), and
compare and swap (CASA). The latter two are atomic load-store
instructions. 

\subsection{Mapping from Instructions to Memory Operation Blocks}
\label{subsec:map_mem_op_block}

\begin{figure}[t!]
    \centering
    \includegraphics[width=\textwidth]{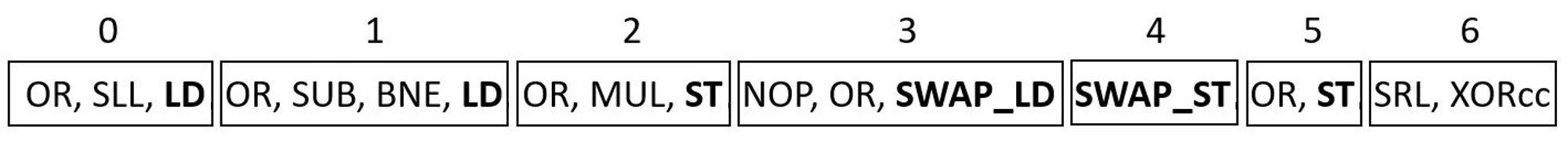}
    \caption{Illustration of memory operation blocks.}
    \label{fig:mem_op_block}
\end{figure}

To bridge the gap between the instruction semantics level and the memory
operation level, we define the concept of \emph{program block} as a list of
instructions where there can be at most one instruction for memory access
(load, store, etc.), and the memory access instruction must be the last
instruction in the list. Intuitively, a block of instructions in the ISA model
corresponds to a memory operation in the memory model, with an exception
discussed below. We illustrate program blocks with the example in
Figure~\ref{fig:mem_op_block}.

Given a list of instructions for the processor to execute, we
identify the memory access instructions (in bold font, such as LD,
ST) and divide the list into several program blocks. In the example
in Figure~\ref{fig:mem_op_block}, there are instructions after the
last memory access instruction, they form a block as well (block 6),
although strictly speaking they are not memory operations. In the
SPARC TSO axiomatic model~\cite{Sindhu1992}, an atomic load-store
instruction is viewed as two memory operations $[L;S]$ where the load
part $L$ and the store part $S$ have to be executed atomically. In
correspondence, we split an atomic load-store instruction, such as
SWAP, into two parts and put them in two consecutive program blocks
(block 3 and 4 in Figure~\ref{fig:mem_op_block}).
We assume that each program block can be uniquely identified.
This gives rise to a mapping $M_{block} = id \Rightarrow block$
from 
an identifier (natural number) to a program block.
The latter is
a tuple $\langle i,p,id\rangle$, where $i$ is a list of instructions, $p$ (natural number) is the processor in charge of executing the code, and $id$ is the identifier of the load part of an atomic load-store instruction (optional).


We distinguish the types of program blocks by the memory operation
involved in it. Program blocks without memory operations are called \emph{non-mem block}, whilst program blocks including memory operations are called \emph{memory operation blocks}.  A \emph{memory operation block} is a 
\emph{load block}
when it has an LD, it is a
\emph{store block} 
when it has an ST. An \emph{atomic load block}
has either SWAP\_LD or CASA\_LD, whereas an \emph{atomic store
block} 
has either SWAP\_ST or CASA\_ST. 

In contrast to the SPARCv8 ISA model, here we lift the processor
execution to be oriented on program blocks, based on the
\emph{program order}. A program order is the order in which a
processor executes instructions~\cite{Sindhu1992}. Since we can identify
program blocks using their 
identifiers
we
define the program order $PO$ for a processor $p$ as a
mapping from $p$ to a list of 
identifiers: $PO = p \Rightarrow id \ list$.

Given a program order $PO$ and a processor $p$, the program blocks in
this program order are related by a \emph{before} relation ``$;$'' as follows:
\begin{definition}[Program Order Before]
\label{def:prog_order_before}
$id_1 \ ;_{PO}^p \ id_2$ iff $id_1$ is before $id_2$ in the list of program block identifiers given by $(PO \ p)$.
\end{definition}



We shall omit the $p$ and/or the $PO$ in the notation of program
order before and write $id_1 \ ; \ id_2$ when the context is
obvious. Only program blocks issued by the same processor can be related by program order. Thus $id_1 \ ; \ id_2$ implicitly identifies a processor.


We divide program execution into two
levels: the processors execute instructions and issue memory
operations in a given program order; the memory executes memory
operations in its own memory order, which will be described in
Section~\ref{sec:mem}.

\subsection{State and Instruction Semantics}
\label{subsec:instr}

The state 
of
a multi-core processor is a 
tuple $\langle ctl, reg, mem, L_{var}, G_{var}, op, undef, next\rangle$, with the following definitions:



$ctl$ are the control registers (per processor), these include Processor State
Register (PSR), which records the current set of registers, whether the
processor is in user mode or supervisor mode, etc.; Program Counter (PC); Next
Program Counter (nPC), among others. $ctl$ is formally defined as a function $ctl = p \Rightarrow C_{reg} \Rightarrow val,$ where $p$ is the processor,
$C_{reg}$ is the control register, $val$ is the value held by the
register (32-bit word).

$reg$ are the general registers (per processor). Formally, $reg = p \Rightarrow r \Rightarrow val,$
where $p$ is the processor, $r$ is the address of the register (32-bit word), and $val$ is the value of the register. SPARC instructions often use three general registers: two source registers, refered to as $rs_1$ and $rs_2$, and a destination register, refered to as $rd$. For instance, the addition instruction takes two values from $rs_1$ and $rs_2$, and store the sum in $rd$. We shall refer to the value $reg\ p\ rx$ of a register $rx$ in processor $p$ as $r[rx]$ when the context of the processor and the state is clear. SPARC fixes the value at register address $0$ to be $0$. So when $rd = 0$, we have $r[rd] = 0$.


A main memory $mem$ is shared by all processors. Similar to the
machine code semantics for x86~\cite{Sewell2010}, we 
focus on
memory access of word (32-bits) only, and we assume that
each memory address points to a word, and data are always
well-aligned. Memory is a (partial) mapping $mem = addr \rightharpoonup val.$

Each processor has a local Boolean variable  $L_{var} = p \Rightarrow bool.$
This Boolean variable is used to record whether
the next instruction should be skipped or not after executing
branching instructions. We refer to this variable as the \emph{annul flag}.

All processors share a global variable $G_{var}$, which is a pair $\langle flag_{atom}, val_{rd}\rangle$, where $flag_{atom}$ is the id of the atomic load block when the processor is executing the corresponding atomic load-store instruction, or is undefined otherwise. $val_{rd}$ stores the value of the general register for destination $rd$ which is used in atomic load-store instructions.



$op$ records a memory operation. Formally, $op = id \Rightarrow \langle op_{addr}, op_{val}\rangle,$
where $id$ is the identifier of the program block for the corresponding memory operation, $op_{addr}$ is the address of the operation, and $op_{val}$ is the value of the operation. For instance, a store operation writes value $op_{val}$ at address $op_{addr}$, whereas a load operation loads value $op_{val}$ from address $op_{addr}$. For a given $id$, $op_{addr}$ and $op_{val}$ are initially undefined. These values are computed during execution of memory blocks.
 

Finally, $undef$ indicates
whether the state is undefined or not, and $next$
gives the index (in the list typically given by $(PO \ p)$) of the next memory
operation to be issued by processor. Formally, $next = p \Rightarrow nat$,
where $p$ is a processor and $nat$ is the index.


To provide consistency w.r.t. the memory model, we split the
semantics of atomic load-store instructions into the load part and
the store part. The processor executes them separately, but the
memory model guarantees that 
their executions are ``atomic''.

We give an example of the formalisation of the CASA instruction
below. The SPARC manual~\cite{sparcv9} specifies the semantics of
CASA as follows, where we adapt the setting from $64$-bit registers
in SPARCv9 to $32$-bit registers in the SPARCv8 model: 
The CASA
instruction compares the register $r[rs2]$ with a memory word pointed
to by the address in $r[rs1]$. If the values are equal, the value in
register $r[rd]$ is swapped with the contents of the memory word
pointed to by the address in $r[rs1]$. If the values are not equal,
the memory location remains unchanged, but the memory word pointed to
by $r[rs1]$ replace the value in $r[rd]$. 
We formalise the core of the load part as below, presented in pseudo-code:

\begin{definition}[CASA Load]
\label{def:casa_load}
$CASA_{load} \ \ addr \ val \equiv $\\
\begin{tabular}{l}
\hspace{10px} \ \textbf{if} $rd \neq 0$ \textbf{then} $val_{rd} \leftarrow r[rd]$; $r[rd] \leftarrow val$; $op_{addr} \leftarrow addr$; $op_{val} \leftarrow val$;\\
\hspace{10px} \ \textbf{else} $val_{rd} \leftarrow r[rd]$; $op_{addr} \leftarrow addr$; $op_{val} \leftarrow val$;
\end{tabular}
\end{definition}

\noindent Given a processor $p$ and the $id$ of a CASA load block, we can
obtain the value $r[rd]$ in processor $p$, and the $\langle op_{addr},
op_{val}\rangle$ pair of the operation. When $rd \neq 0$, we store $r[rd]$ in
the temporary global variable $val_{rd}$, and write $val$ into $rd$. We then
store $addr$ and $val$ in $op_{addr}$ and $op_{val}$ respectively. When $rd =
0$, we do not have to write the $rd$ register because its value must be $0$.
In this definition, $addr$ is obtained from $r[rs_1]$, and $val$ (the value at address $addr$) is obtained from Axiom Value of the TSO model which is described in Section~\ref{subsec:axiom_model}. 
The store part is given below:


\begin{definition}[CASA Store]
\label{def:casa_store}
$CASA_{store} \ \ addr \equiv $\\
\begin{tabular}{l}
\hspace{10px} \ \textbf{if} $r[rs_2] = op_{val}$ \textbf{then}  $op_{addr} \leftarrow addr$; $op_{val} \leftarrow val_{rd}$;
\end{tabular}
\end{definition}

\noindent 
We check if $r[rs_2]$ has the same value as $op_{val}$, which corresponds to $val$ in the load part. If this is the case, we then update $op_{addr}$ and $op_{val}$ with $addr$ and $val_{rd}$ respectively, where $addr$ is the same as the address in the load part. 
Note that instruction semantics is only for
processor execution, which does \emph{not} update the memory. Memory write occurs in the store operation defined in the operational semantics of the TSO model, which is introduced in Section~\ref{subsec:op_model}. 

\subsection{Processor Execution}
\label{subsec:proc_exe}

Processor execution includes three stages: fetch, decode, and
dispatch. Since this model is built for analysing memory operations,
we assume that there is a given program order from which we fetch the
instructions. This is similar to the concept of ``run skeletons'' in
the x86 weak memory models~\cite{Sarkar2009}. Decoding facilities
are provided by the SPARCv8 ISA model~\cite{hou2016}. Dispatching and
executing the instructions require more care because we will be
executing blocks (lists) of instructions at a time. For simplicity we
only discuss three interfaces in prose here.

\begin{definition}[$exe$]
Given a processor $p$, program order $PO$, program block map $M_{block}$,
a memory operation block identified by $id$,
and $state$,
the function 
\vspace{-3px}
\begin{center}
$exe \ \ p \ PO \ M_{block} \ id \ state$
\end{center}
\vspace{-3px}
executes the program blocks in the list given by $(PO \ p)$ from the
position 
given by $(next \ p)$  
to the position 
of $id$
(inclusive). The function returns the state after the above execution. We may simplify the above and write $exe_{id} \ state$.
\end{definition}

The function $exe$ is used for executing store blocks, atomic store blocks, and
non-mem blocks. Load and atomic load blocks require more execution steps. We
define the following functions to handle them, assuming the same parameters:

\begin{definition}[$exe^{pre}$]
The function    
\vspace{-3px}
\begin{center}
$exe^{pre} \ \ p \ PO \ M_{block} \ id \ state$
\end{center}
\vspace{-3px}
executes until the instruction before the last one in the block $id$. The function returns the state after the above execution. We may simplify the above and write $exe^{pre}_{id} \ state$.
\end{definition}

Take Fig.~\ref{fig:mem_op_block} for example, if $i = 3$, then $exe^{pre}_3$
executes up to the OR instruction and then stops without executing the SWAP\_LD
instruction.

\begin{definition}[$exe^{last}$]
The function
\vspace{-3px}
\begin{center}
$exe^{last} \ \ p \ PO \ M_{block} \ i \ val \ state$
\end{center}
\vspace{-3px}
which takes an additional 32-bit word value $val$ as input, executes the last instruction in the block $id$. The function returns the state after the above execution. We may simplify the above and write $exe^{last}_{id} \ val \ state$. 
\end{definition}


The $exe^{last}_{id} \ val$ function essentially executes the load (or atomic
load) instruction by loading the value $val$ from memory. Again, take
Fig.~\ref{fig:mem_op_block} for example, when $i = 3$, $exe^{last}_3 \ val$
executes the SWAP\_LD instruction. Note that we do not need the extra input
$val$ for executing store instructions because both the address and the value
for a store can be pre-computed from the instruction code. For load
instructions, however, only the address can be pre-computed from instruction
code. We need to execute until the instruction before the load instruction,
then invoke the memory model to determine the value $val$ to be loaded, which
is why we need two steps when executing a load (or atomic load) block.
 

In this setup, when executing a memory load operation, all previous
memory operations in the program order have been executed, and their
corresponding addresses ($op_{addr}$) and values ($op_{val}$) have been
updated in the state. This allows us to directly use the SPARC TSO
Axiom Value (cf. Section~\ref{subsec:axiom_model}) to obtain the
value of the load operation. 

\section{SPARC TSO Memory Model}
\label{sec:mem}


Details of the SPARC TSO model can be
found in \cite{Sindhu1992,sparcv8}.
This section formalises the 
axiomatic model in Isabelle/HOL. More importantly, we give a novel
operational model, and show that the operational model corresponds to
the axiomatic model. 


\subsection{Axiomatic TSO Model}
\label{subsec:axiom_model}

The complete semantics of TSO are captured by six
axioms~\cite{Sindhu1992,sparcv8}, which specify the \emph{ordering}
of memory operations. The semantics of loads and stores to I/O
addresses are implementation-dependent and are not covered by the TSO
model. The SPARCv8 manual only specifies that loads and stores to I/O
addresses must be strongly ordered among themselves. We adapt these
axioms to our abstract SPARC ISA model and formalise them in
Isabelle/HOL. Similar to the x86-TSO model~\cite{Owens2009}, we focus
on data memory, thus our memory model does not consider instruction
fetch and flush.

Besides the \emph{program order before} relation (cf. Definition~\ref{def:prog_order_before}),
the axiomatic model also
relies on a \emph{before} relation over operations but in \emph{memory order},
which
is the order that the
memory executes load and store operations. 
Given a partial/final memory
execution represented by a sequence $x$ of $id$s, the before
relation over two operations $id_1$ and $id_2$ in memory order is
defined below as a partial function from the pair to $bool$, where we write $id \in x$ when $id$ is in the sequence $x$:

\begin{definition}[Memory Order Before]
\label{def:mem_order_before}
$id_1 \ <_{x} \ id_2 \equiv$\\
\begin{tabular}{l} 
\hspace{10px} \ \textbf{if} $(id_1\in x) \ \land \ (id_2 \in x)$ \textbf{then}\\
\hspace{10px} \ \ \ \textbf{if} $id_1$ is before $id_2$ in $x$ \textbf{then} $true$ \textbf{else} $false$\\
\hspace{10px} \ \textbf{else if} $id_1\in x$ \textbf{then} $true$ \textbf{else if} $id_2\in x$ \textbf{then} $false$ \textbf{else} $undefined$
\end{tabular}
\end{definition}

\noindent 
We
may loosely refer to a memory order by the corresponding partial/final memory
execution sequence $x$. We may write $id_1 \ < \ id_2$ when the context
is clear. Note that any memory operation \emph{id} in the sequence of executed operations \emph{x} has been already executed by the processor and thus $op_{addr}\ id$ in the current state is defined.

The axiom \emph{Order} states that in a final execution sequence $x$,
every pair $id, id'$ of store operations are related by $<_{x}$. 
This axiom is
formalised as below:

\begin{definition}[Axiom Order]
\label{def:ax_order}
$order \ id \ id' \ x \ M_{block} \equiv$\\
\textbf{If} both $(M_{block} \ id)$ and $(M_{block} \ id')$ are either a store or an atomic store block, and both $id$ and $id'$ are in $x$, and $id\neq id'$, \textbf{then} either $(id \ <_{x} \ id')$ or $(id' <_{x} id)$.
\end{definition}


The axiom \emph{Atomicity} ensures that for an atomic load-store instruction,
the load part $id_l$ is executed by the memory before the store part $id_s$, and there
can be no other store operations executed between $id_l$ and
$id_s$. 
\begin{definition}[Axiom Atomicity]
\label{def:ax_atom}
$atomicity \ id_l \ id_s \ PO \ x \ M_{block} \equiv$\\ 
\hspace{10px} \textbf{If} \ $id_l$ and $id_s$ are from the same instruction instance, and $(id_l \ ; \ id_s)$, and $(M_{block} \ id_l)$ is an atomic load block, and $(M_{block} \ id_s)$ is an atomic store block, \textbf{then} $id_l \ <_{x} \ id_s$, and for all store or atomic store block $(M_{block} \ id)$,  if $id \in x$ and $id \neq id_s$, then either $id \ <_{x} \ id_l$ or $id_s \ <_{x} \ id$.
\end{definition}

The axiom \emph{Termination} states that all store operations eventually
terminate. 
We capture this by ensuring that after the execution is
completed, every store operation $id$ that appears in the program list of some processor is in the sequence $x$ of executed
operations. We formalise this axiom as follows:

\begin{definition}[Axiom Termination]
\label{def:ax_term}
$termination \ id \ PO \ x \ M_{block} \equiv $\\
\textbf{If} there exists a processor $p$ such that $id \in (PO \ p)$, and $(M_{block} \ id)$ is a store or atomic store block, \textbf{then} $id \in x$.
\end{definition}

The axiom \emph{Value} states that the value of a load operation
$id$ issued by processor $p$ at address $addr$ is the value written
by the most recent store to that address. The most recent store at
$addr$ could be: (1) the most recent store issued by processor $p$, 
or (2) the most recent store (issued by any processor) executed by
the memory. 

\begin{definition}[Axiom Value]
\label{def:ax_value}
$value \ p \ id \ addr \ PO \ x \ M_{block} \ state \equiv$\\ 
Let $Max_{<}$ denote a function that outputs the last element in the order defined by $<$ (memory order before) in a set of $id$s.

\begin{quote}
$Max_{<}$ $(\{id' \mid id' <_x id$, and $(M_{block} \ id')$ is a store or atomic store block, and $addr$ is equal to $op_{addr}$ of $id' \} \cup \{id' \mid id' \ ; \ id$ and $(M_{block} \ id')$ is a store or atomic store block, and $addr$ is equal to $op_{addr}$ of $id'\})$,
\end{quote}
the value to be loaded is $op_{val}$ of the output of $Max_{<}$.
\end{definition}

\noindent Intuitively, the output of $Max_{<}$ is the last element in the order given by $<$
from two sets of block ids: The first set includes all the store operations
that are before $id$ in the memory order $x$ and write values at address
$addr$. The second set includes all the store operations that are before $id$
in the program order (given by $(PO \ p)$) and write values at address $addr$.
Therefore  $Max_{<}$ returns the most recent store operation at address $addr$ in
memory order. We write $Lval_{id}$ to denote the value to be loaded for
operation $id$ based on Axiom Value.




The axiom \emph{LoadOp} requires that any operation $id'$ issued
after a load $id$ in the program order must be executed by the
memory after $id$. This is formalised as below:

\begin{definition}[Axiom LoadOp]
\label{def:ax_loadop}
$loadop \ id \ id' \ PO \ x \ M_{block} \equiv$\\ 
\textbf{If} $(M_{block} \ id)$ is a load or atomic load block, and $id \ ; \ id'$, \textbf{then} $id \ <_{x} \ id'$.
\end{definition}

The axiom \emph{StoreStore} states that if a store operation
$id$ is before another store operation $id'$ in the program
order, then $id$ is before $id'$ in the memory order.

\begin{definition}[Axiom StoreStore]
\label{def:ax_storestore}
$storestore \ id \ id' \ PO \ x \ M_{block} \equiv$\\
\textbf{If} $(M_{block} \ id)$ and $(M_{block} \ id')$ are store or atomic store blocks, $id \ ; \ id'$, \textbf{then} $id <_x id' $.
\end{definition}

\begin{figure}[t!]
\centering
\begin{tabular}{c}
\AxiomC{$type_{id} = ld$}
\AxiomC{$\forall id'. \ ((id' \ ; \ id) \ \land \ 
type_{id'} \in \{ld,ald\} \longrightarrow \ id'\in x)$}
\alwaysSingleLine
\RightLabel{\scriptsize $load$}
\BinaryInfC{$x,s \leadsto x@[id],(exe^{last}_{id} \ Lval_{id}  \ (exe^{pre}_{id}  \ s))$}
\DisplayProof\\[30px]
\AxiomC{$type_{id} = st$}
\alwaysNoLine
\AxiomC{$flag_{atom} = undefined$}
\BinaryInfC{$\forall id'. ((id' \ ; \ id) \ \land \ type_{id'} \in \{ld, ald, st, ast\} \longrightarrow id'\in x)$}
\alwaysSingleLine
\RightLabel{\scriptsize $store$}
\UnaryInfC{$x,s \leadsto x@[id],(W_{mem} \ id \ (exe_{id} \ s))$}
\DisplayProof\\[30px]
\AxiomC{$type_{id} = ald$}
\alwaysNoLine
\AxiomC{$flag_{atom} = undefined$}
\BinaryInfC{$\forall id'. ((id' \ ; \ id) \ \land \ type_{id'} \in \{ld, ald, st, ast\} \longrightarrow id'\in x)$}
\RightLabel{\scriptsize $atom\_load$}
\alwaysSingleLine
\UnaryInfC{$x,s \leadsto x@[id],(flag^{set}_{atom} \ id \ (exe^{last}_{id} \ Lval_{id}  \ (exe^{pre}_{id}  \ s)))$} 
\DisplayProof\\[30px]
\AxiomC{$type_{id} = ast$}
\alwaysNoLine
\AxiomC{$flag_{atom} = id'$}
\AxiomC{$atom_{pair} \ id = id'$}
\TrinaryInfC{$\forall id''. ((id'' \ ; \ id) \ \land \ type_{id''} \in \{ld, ald, st, ast\} \longrightarrow id''\in x)$} 
\RightLabel{\scriptsize $atom\_store$}
\alwaysSingleLine
\UnaryInfC{$x,s \leadsto x@[id],(W_{mem} \ id \ (flag^{set}_{atom} \ undef \ (exe_{id} \ s)))$}
\DisplayProof
\end{tabular}
\caption{Rules for the operational TSO model.}
\label{fig:op_tso_model}
\end{figure}

\subsection{Operational TSO Model}
\label{subsec:op_model}

Compared with other operational memory
models such as the x86-TSO model~\cite{Sewell2010}, our ISA model
enables us to develop a more abstract operational memory model
without using concrete modules such as store buffer, which
effectively buffers the address and value 
of most recent store
operations. This alleviates the burden of modelling complicated
operations and interactions between the processor and the store
buffer, and results in a simple and elegant operational memory model.
Our operational TSO model is defined via inference rules. 
An operation takes the form  $x,s \leadsto x',s'$  where 
$x$ and $s$ are respectively the partial execution sequence and state \emph{before} the operation, and $x'$ and $s'$ are respectively the partial execution sequence and state \emph{after} the operation.

We shall use the following notations: 
We write $type_{id}$ to denote the type of the memory
operation block ($M_{block} \ id$). We use the following abbreviations for
memory operation block types: ld (load), ald (atomic load), st (store), ast
(atomic store), non (non-mem). We write $x@x'$ for the concatenation of two
sequences $x$ and $x'$. We write $W_{mem} \ id \ s$ for memory commit (write) of operation $id$ in state $s$. We define the operation $flag^{set}_{atom} \ id \ s$ to set the atomic flag $flag_{atom}$ to $id$ in state $s$. This operation returns a new state. We write $flag^{set}_{atom} \ undef \ s$ to set the flag to undefined. When the operation $id$ is an atomic store operation, the function $atom_{pair} \ id$ returns the operation $id'$ such that $id'$ is the corresponding atomic load operation of the same instruction. This function is otherwise undefined.

The operational TSO model consists of four rules, which are given in
Figure~\ref{fig:op_tso_model}. 
The first rule for load
operations has two premises: (1) the type of the operation $id$ is
load; (2) every load operation before $id$ in the program
order has been executed by the memory. The operation first executes ($exe^{pre}_{id}$) all instructions in the program order before the last instruction (which must be the load instruction) in the block $id$, then uses Axiom Load ($Lval_{id}$) to determine the value to be loaded, and finally executes ($exe^{last}_{id}$) the load instruction.

The rule for store operations requires that $flag_{atom}$ in state $s$ must be
undefined. That is, the memory is not in the middle of executing an atomic
load-store operation. Also, the rule requires that every load \emph{or store}
operation before $opid$ in the program order has been executed by the memory.
Combining the last premise of $load$, $atom\_load$, and $atom\_store$
respectively, these requirements ensure that axioms LoadOp and StoreStore are
respected in execution. For instance, it is possible that a store is issued (by
a processor) before a load but is executed (by memory) after the load; but it
is not possible that a load is issued before a store but executed after the
store. The store operation's final step is to commit the store operation $id$
in memory. This step fetches the value $op_{val}$ and address $op_{addr}$ of
the operation $id$ from the state, and writes the value at the address in the
memory.

The premises for the rule $atom\_load$ can be read similarly. The final step of
the $atom\_load$ operation sets $flag_{atom}$ to $id$, where $id$ is the atomic
load operation. Accordingly, the rule $atom\_store$ requires that the memory
has executed the atomic load part $id'$, but has not executed the store part.
The rule $atom\_store$ also ensures that the $atomic_{pair}$ of the store part
$id$ is indeed $id'$. The operation eventually sets the $flag_{atom}$ back to
undefined and commits the operation in the memory. The premises with regard to
$flag_{atom}$ and $atomic_{pair}$ ensure that axiom Atomicity holds in
execution.

In addition to the rules for memory operations, to obtain the final result of processor execution, we may need the rule $non\_mem$:
\begin{center}
\AxiomC{$type_{id} = non$}
\alwaysNoLine
\AxiomC{$\forall id'. ((id' \ ; \ id) \ \land \ type_{id'} \in \{ld, ald, st, ast\} \longrightarrow id'\in x)$}
\RightLabel{\scriptsize $non\_mem$}
\alwaysSingleLine
\BinaryInfC{$x,s \leadsto x@[id],(exe_{id} \ s)$}
\DisplayProof
\end{center}
This rule executes the block after the last memory operation (e.g., block
6 in Figure~\ref{fig:mem_op_block}), if there is any. 
This rule is not related to the
memory model because it does not involve memory operations. It
plays no roles in the proofs in the remainder of this section.

\subsection{Soundness and completeness of the operational model}

We are now ready to present the main results of this work: the operational TSO model is sound and complete w.r.t. the TSO axioms.
The previous subsection has briefly discussed that the design of
operational rules respects the axioms such as LoadOp, StoreStore, and
Atomicity. Axiom Value trivially holds in the operational model
because the rule $load$ directly uses axiom Value to obtain load
result. 
Axiom Termination is satisfied by the construction of the execution witness
sequences, because the $x$ part of the final witness is guaranteed to contain all the store operations, which means that the execution of these operations have been completed by the memory.
Axiom Order holds because all the executed store operations are recorded in a list, which means every pair of them are ordered.
The formal proof of the correspondence of
the axiomatic model and the operational model is rather complicated,
and here we only discuss the results. Interested readers can check the Isabelle/HOL formalization and proofs\footnote{Appendix with proofs is at http://securify.sce.ntu.edu.sg/MicroVer/SparcTSO/appendix.pdf} for more details. 

\begin{theorem}[Soundness]
Every memory operation sequence generated by the operational model satisfies the axioms in the axiomatic model. 
\end{theorem}

\begin{theorem}[Completeness]
Every memory operation sequence that satisfies the axioms in the axiomatic model can be generated by the operational model. 
\end{theorem}
\section{Case Studies}
\label{sec:case}


With the above work, we can now formally reason about concurrent machine code. The axiomatic model can be used to reason about the order of memory operations, while the operational model is better at reasoning about properties of the execution flow. 
We run two case studies drawn from examples in the SPARCv9
manual~\cite{sparcv9}. 
We may use the term
\emph{process} and \emph{processor} interchangeably. See Owen's work~\cite{Owens2010} for a semantic foundation for reasoning
about programs in TSO-like relaxed memory models.

\subsection{Indirection Through Processors}


\begin{table}[t!]
		\centering
		\begin{tabular}{|c|c|l|}
			\hline
			Processor & $op\_id$ & Instruction\\
			\hline
			1 & 0 & $OR \ \ \ \ \%g0, 1, \%r4$\\
			& & $OR \ \ \ \ \%g0, 1, \%r5$\\
			& & $ST \ \ \ \ \%r5,[\%g0 + \%r4]$\\
			\cline{2-3}  
			& 1 & $OR \ \ \ \ \%g0, 1,\%r5$\\
			& & $OR \ \ \ \ \%g0, 2,\%r4$\\
			& & $ST \ \ \ \ \%r5,[\%g0 + \%r4]$\\
			\hline
			2 & 2 & $OR \ \ \ \ \%g0, 2, \%r4$\\
			& & $LD \ \ \ \ [\%g0 + \%r4], \%r1$\\
			\cline{2-3} 
			& 3 & $OR \ \ \ \ \%g0, 3, \%r4$\\
			& & $ST \ \ \ \ \%r1, [\%g0 + \%r4]$\\
			\hline
			3 & 4 & $OR \ \ \ \ \%g0, 3, \%r4$\\
			& & $LD \ \ \ \ [\%g0 + \%r4], \%r1$\\
			\cline{2-3} 
			& 5 & $OR \ \ \ \ \%g0, 1, \%r4$\\
			& & $LD \ \ \ \ [\%g0 + \%r4], \%r2$\\
			\hline
		\end{tabular}
		\vspace{10px}
		\caption{``Indirection Through Processors".}
		\label{tab:ex_itp}
	\end{table}

The ``Indirection Through Processors" program is taken from Figure 46
of the SPARCv9 manual~\cite{sparcv9}. This example intends to reflect
the TSO property that causal update relations are preserved.
The original program involves three processors,
each processor issues two memory operations. A memory operation is given in an  ``instruction-like" style, e.g.,  $st \ \ \ \ \ \#1,[A]$ means that the value $1$ is stored into address $A$ of the memory. Unfortunately in real SPARC store instructions, 
the value to be stored and the value of the memory address must be taken from registers, so we need to add a few instructions to initialise the registers for this example to work. Our formalised ``Indirection Through Processors" example is shown in Table~\ref{tab:ex_itp}. 
The global register $\%g0$ in SPARC always contains $0$. The first
instruction in block $0$ adds $0$ and $1$, and puts the result in
register $\%r4$. The $ST$ in block $0$ thus stores $1$ at memory
address $1$. The $ST$ in block $1$ stores $1$ at address $2$. The
$LD$ in block $2$ loads the value at address $2$ to register $\%r1$.
Block $3$ then stores the value in $\%r1$ at address $3$. Finally,
processor $3$ loads the values at addresses $3$ and $1$ to registers
$\%r1$ and $\%r2$. 

\paragraph{Reasoning about memory operation order.}
It is intuitive to use the axiomatic TSO model to reason about the
order of memory operations. For the program in
Table~\ref{tab:ex_itp}, the SPARCv9 manual gives some example
sequences of memory operations allowed under TSO, and an example
sequence that is not allowed under TSO: $x =
[1,2,3,4,5,0]$. This is because $(0 \ ; \ 1)$ must hold in the program order given by Table~\ref{tab:ex_itp}, and the above sequence implies
that $\lnot (0 \ < \ 1 = true)$ in the memory order, which falsifies the axiom
StoreStore.

Alternatively, the completeness of the operational TSO model enables
us to use the operational model to reason about the possible next
step operations. The above reasoning can be confirmed by our
operational model in the lemma below:

\begin{lemma} 
  \label{lem:case1}
$init \longrightarrow \lnot ([],s \leadsto [1],s')$
\end{lemma}

Lemma~\ref{lem:case1} states that given a partial execution sequence which contains only an
initialisation step $init$ where memory addresses are set to $undefined$ and registers
are set of $0$, memory operation block $1$ in Table~\ref{tab:ex_itp} cannot be
the first operation to be executed. 



\paragraph{Reasoning about execution result.}
Besides eliminating illegal executions, one can also use our
operational model to reason about the results of legal
executions. For instance, the SPARCv9 manual lists the sequence
$x' = [0,1,2,3,4,5]$ as a legal execution under TSO. For
simplicity, here we only show that after a partial execution $[0,1,2]$, the
register $\%r1$ of processor $2$ has value $1$, which is stored to
address $2$ by processor 1 previously. This shows that a processor can observe
the memory updates made by other processors. This is formalised in
the following lemma:

\begin{lemma} 
  $[0,1],s_2 \leadsto [0,1,2],s_3 \ \longrightarrow (reg \ s_3) \ 2 \ 1 = 1$
\end{lemma}


\noindent The right hand side of the implication means that in state $s_3$,
the general register $1$ of processor $2$ contains value $1$. The proof for
execution results usually involves a ``simulation'' of the execution using the
abstract ISA model and the operational TSO model. For this example, we start
from the initial witness, and prove a series of lemmas about the execution
witnesses $([0],s_1), ([0,1],s_2), ([0,1,2], s_3)$ for the intermediate
execution steps. It is straightforward to complete this series of proofs and obtain
the result of a final execution.

\subsection{Spin Lock with Compare and Swap}

\begin{figure}[t!]
\centering
\begin{subfigure}[b]{0.4\textwidth}
\centering
\begin{tabular}{l}
\textbf{Lock}$(lock, proc\_id)$ \\
retry: \\
\ \ mov $\ \ [proc\_id], \%l0$\\
\ \ cas $\ \ [lock], \%g0, \%l0$\\
\ \ tst $\ \ \%l0$\\
\ \ be $\ \ $ out\\
\ \ nop\\
loop: \\
\ \  ld $\ \ [lock], \%l0$\\
\ \  tst $\ \ \%l0$\\
\ \  bne $\ \ $ loop\\
\ \  nop\\
\ \  ba,a $\ \ $ retry\\
out:  \\
\ \ code in critical region \\[5px]
\textbf{Unlock}$(lock)$ \\
\ \  st $\ \ \%g0, [lock]$\\
\end{tabular}
\caption{Spin lock using CASA.}
\label{fig:ex_spin_a}
\end{subfigure}
\begin{subfigure}[b]{0.5\textwidth}
\centering
\begin{tabular}{|c|c|l|}
\hline
Processor & $op\_id$ & Instruction\\
\hline
1 & 0 & $OR \ \ \ \ \%g0, 1, \%r16$\\
  & & $OR \ \ \ \ \%g0, 1, \%r1$\\
  & & $CASA\_LD \ \ \ \ [\%r1],\%g0,\%r16$\\
\cline{2-3}  
  & 1 & $CASA\_ST \ \ \ \ [\%r1],\%g0,\%r16$\\
\cline{2-3}
 & 5 & $ORcc \ \ \ \ \%g0, \%r16, \%g0$\\
 & & $BE \ \ \ \ 28$\\
 & & $NOP$\\
\hline
2 & 2 & $OR \ \ \ \ \%g0, 1, \%r16$\\
  & & $OR \ \ \ \ \%g0, 1, \%r1$\\
  & & $CASA\_LD \ \ \ \ [\%r1],\%g0,\%r16$\\
\cline{2-3} 
  & 3 & $CASA\_ST \ \ \ \ [\%r1],\%g0,\%r16$\\
\cline{2-3}
 & 6 & $ORcc \ \ \ \ \%g0, \%r16, \%g0$\\
 & & $BE \ \ \ \  28$\\
 & & $NOP$\\  
\hline
3 & 4 & $OR \ \ \ \ \%g0, 1, \%r4$\\
 & & $ST \ \ \ \ \%g0, [\%g0 + \%r4]$\\
\hline
\end{tabular}
\vspace{10px}
\caption{A fragment of formalised spin lock code.}
\label{fig:ex_spin_b}
\end{subfigure}
\caption{The spin lock example.}
\label{fig:ex_spin}
\end{figure}


Section J.6 of the SPARCv9 manual~\cite{sparcv9} gives an example of spin
lock implemented using the CASA instruction
, the code is shown in
Figure~\ref{fig:ex_spin_a}. 
Note that the code in Figure~\ref{fig:ex_spin_a} is in
\emph{synthetic instruction} format. SPARCv8/v9 manual provides a
straightforward mapping from this format to \emph{SPARC instruction}
format, which is what our ISA model supports. For instance, in the
\emph{retry} fragment, the first instruction $mov$ corresponds to an
$OR$, which adds the ID $proc\_id$ of the current process and $0$,
and stores the result to register $\%l0$, which corresponds to
register $\%r16$. After executing this line, $\%l0$ ($\%r16$)
contains the ID of the current process. The second line is the CASA
instruction. It checks whether the memory value at address $lock$ is equal
to the value at $\%g0$ (which must be $0$), and swaps the value at
address $lock$ and the value at register $\%l0$ when the above check
is positive. Otherwise, the value at address $lock$ is stored at
register $\%l0$. Therefore, when no processes hold the lock, the
value at address $lock$ is $0$, and after executing the second line,
$\%l0$ ($\%r16$) will have $0$ and address $lock$ will contain the ID
of the current process. On the other hand, when the lock is held by
another process, after executing CASA, the memory address $lock$ is
unchanged, and $\%l0$ contains the ID of the process that holds the
lock. The code $tst \ \ \%l0$ corresponds to an $ORcc$, which checks
if $\%l0$ is equal to $0$. If it is, then the program branches to
$out$, and starts to execute in the critical region. Otherwise, the
program goes to $loop$ and keeps reading the address $lock$ until it
contains a $0$.


We give the fragment of instructions before entering the critical region in
Figure~\ref{fig:ex_spin_b}, and consider a concrete situation where two
processes (processors) $1$ and $2$ are competing to get the lock, and
process $3$ initialises the lock to $0$. 
Assume that process $3$
executes operation $4$ first for initialisation, also assume without
of loss generality that operation $0$ of process $1$ is executed by
the memory earlier than process $2$'s operations, we show that
process $1$ will enter the critical region. The case where operation
$2$ of process $2$ is executed earlier by the memory is symmetric. In
this example, we set the address of 
  critical region as $28
\ll 2 = 112$ relative to the address of the branch instruction $BE$,
where $\ll$ is \emph{sign extended shift to the left}.

The proof uses a mixture of the techniques in the previous subsection
to obtain valid memory operation sequences and reason about the
results. We omit the intermediate steps and show the final lemma
below:


\begin{lemma}
    $[4,0,1,2,3,5],s_6 \leadsto [4,0,1,2,3,5,6],s_7 \longrightarrow (ctl \ s_7) \ 1 \ nPC = (ctl \ s_7) \ 1 \ PC + 112 \ \land (ctl \ s_7) \ 2 \ nPC = (ctl \ s_7) \ 2 \ PC + 4$
\end{lemma}

\noindent The right hand side of the implication shows that the $nPC$
(next program counter) of processor $1$ is the entry point of the
critical region, while the $nPC$ of processor $2$ points to $NOP$,
after which will lead processor $2$ to the loop in Figure~\ref{fig:ex_spin_a}.
\section{Conclusion and Future Work}
\label{sec:conc}

This paper gives an abstraction of the SPARCv8 ISA model in
Isabelle/HOL~\cite{hou2016}. The new model is suitable for formal
modelling and verification at the memory operation level. We also
extend the ISA model with semantics for the SPARCv9 instruction
Compare and Swap, which is useful in concurrent programs. The more
abstract ISA model splits the semantics for atomic load-store
instructions into two parts: the load part and the store part, which
correspond to the operations in the memory model.

On top of the abstract ISA model, we formalise the SPARC TSO
axiomatic memory model in Isabelle/HOL. This model is useful for
reasoning about the order of memory operations. We also give a novel
operational TSO memory model as a system that consists of four rules.
We show that the operational TSO model is sound and complete with
respect to the axiomatic model. Finally, we demonstrate the use of
our memory models with two examples in the SPARCv9 manual.

All the models and proofs in this paper are formalised in
Isabelle/HOL. The abstract SPARC ISA model measures 1960 lines of
code, the two memory models and the soundness and completeness proofs
constitute 4753 lines of code, the case studies take up 1750 lines of
code.

One of our next steps is to generate executable code from our operational TSO
model and conduct experiment against real hardware. One can view this as a
``validation'' step. However, our understanding of the SPARC TSO model is that
the TSO axiomatic model came as a part of the SPARCv8 manual before the
implementation of actual hardware, thus the TSO axiomatic model should be
seen as a standard that the hardware must comply rather than the other way
around. Therefore a better validation would be to show that our formalisation
of the TSO axiomatic model is consistent with the definitions in the SPARCv8
manual, which is easy to verify.

Our current on-going work is about developing a Hoare-style
logic for SPARC machine code. The current framework, which includes
the abstract ISA model and the memory models, provides the
foundation for the verification of concurrent machine code. However,
if a program involves a complex control-flow with branches and loops,
it is tedious to use the current models to reason about the
program. A Hoare-style logic is much desired to make the reasoning
task easier. 
We envision that this new work will make it easier to prove
properties such as reachability, safety, and non-interference.

\bibliographystyle{splncs03}
\bibliography{main}


\end{document}


\title{A formalisation of the SPARC TSO memory model for concurrent machine code (Appendix)}  

\titlerunning{abbreviated title}  
%
\author{Zhe Hou\inst{2} \and David San\'{a}n\inst{1} \and Alwen Tiu\inst{3}  \and Yang Liu\inst{1} \and Jin Song Dong\inst{2}}
\authorrunning{Hou et al.} 
%
%
\institute{School of Computer Science and Engineering, Nanyang Technological University, Singapore
  \and
  Institute for Integrated and Intelligent Systems, Griffith University, Australia
  \and
  Research School of Computer Science, The Australian National University, Australia}

\maketitle



  \appendix
 \section{Appendix}
\label{sec:app}

\subsection{Soundness of the Operational TSO Model}
\label{subsec:op_sound}

This subsection shows that every final memory execution given by the
operational TSO model satisfies the TSO axioms.

We define the execution constructed by the operational TSO model in
terms of \emph{execution witnesses}. An execution witness is a triple
$(xseq, rops, s)$ where $xseq$ is a list of $op\_id$s that have been
executed by the memory, $rops$ is the set of $op\_id$s that are
remaining to be executed, and $s$ is the current state. In the
sequel, given an execution witness $exe$, we will use subscripts
$exe_{xseq}$, $exe_{rops}$, and $exe_s$ to denote the corresponding
parts of the witness. Transition between execution witnesses is
defined as below:

\begin{definition}[Execution Witness Transition]
\label{def:exe_wit_trans}
$po \ pbm \vdash_t \ (xseq,rops,s) \leadsto (xseq',rops',s') \equiv $\\
\begin{tabular}{l}
\hspace{10px} \ $\exists opid.\ opid \in rops \ \land \ po \ pbm \ opid \vdash_m xseq,s \leadsto xseq',s' \ \land \ rops' = rops - \{opid\}$
\end{tabular}
\end{definition}

We assume that a \emph{valid program} given by a program order $po$
and program block mapping $pbm$ should satisfy the following
conditions:

\begin{definition}[Valid Program]
\label{def:valid_prog}
$valid\_program \ po \ pbm \equiv $\\
\begin{tabular}{l}
\hspace{10px} \ $(\forall opid \ opid' \ opid''.\  (atom\_pair\_id \ (pbm \ opid) = Some \ opid' \ \land \ opid \neq opid'') \ \longrightarrow$\\
\hspace{10px} \ \ \ $(atom\_pair\_id \ (pbm \ opid'') \neq Some \ opid')) \ \land$\\
\hspace{10px} \ $(\forall opid \ p.\ (member \ (po \ p) \ opid \ \land \ op\_type \ (pbm \ opid) = ald\_block) \ \longrightarrow$\\ 
\hspace{10px} \ \ \ $(\exists opid'.\  (opid \ ;_{po}^p \ opid') \ \land \  op\_type \ (pbm\  opid') = ast\_block  \ \land $\\
\hspace{10px} \ \ \ $atom\_pair\_id \ (pbm \ opid') = Some \ opid)) \ \land$\\
\hspace{10px} \ $(\forall opid' \ p.\ (member \ (po \ p) \ opid' \ \land \ op\_type \ (pbm \ opid') = ast\_block) \ \longrightarrow$\\  
\hspace{10px} \ \ \ $(\exists opid.\ (opid \ ;_{po}^p \ opid') \ \land \ op\_type \ (pbm \ opid) = ald\_block \ \land$\\
\hspace{10px} \ \ \ $atom\_pair\_id \ (pbm \ opid') = Some \ opid)) \ \land $\\
\hspace{10px} \ $(\forall p.\ non\_repeat\_list \ (po \ p)) \ \land$\\
\hspace{10px} \ $(\forall opid \ opid' \ p.\ (opid \ ;_{po}^p \ opid') \ \longrightarrow$\\ 
\hspace{10px} \ \ \ $(proc\_of\_op \ opid \ pbm = proc\_of\_op \ opid' \ pbm
   \ \land \ p = proc\_of\_op \ opid \ pbm)) \ \land$\\
\hspace{10px} \ $(\forall opid \ p.\ (member \ (po \ p) \ opid) \ \longrightarrow \ (proc\_of\_op \ opid \ pbm = p))$
\end{tabular}
\end{definition}

The first conjunct states that for any two distinct operations $opid$ and
$opid''$, their atomic pair (the corresponding atomic load part)
cannot be the same. So given an atomic store operation, we can
identify a unique atomic load operation. The second conjunct states that given an atomic load operation $opid$ which is in the program list given by $(po \ p)$, there exists a corresponding atomic store operation $opid'$ that pairs with $opid$ in the same program list. The third conjunct states the other way around. The fourth conjunct requires that every program list $(po \ p)$ does not have repeating elements. That is, every instance of $op\_id$ is unique. The fifth conjunct requires that if $opid$ is before $opid'$ in the program order, then they are executed by the same processor. The last condition is that the function $proc\_of\_op$ correctly gives the processor that executes the operation.

A valid partial execution witness list $exe\_list$ constructed by the
rules of the operational TSO model is defined as follows:

\begin{definition}[Valid Memory Execution Sequence]
\label{def:valid_mem_exe_seq}
$valid\_mem\_exe\_seq \ po \ pbm \ exe\_list \equiv $\\
\begin{tabular}{l}
\hspace{10px} \ $valid\_program \ po \ pbm \ \land \ length \ exe\_list > 0 \ \land \ (hd \ exe\_list)_{xseq} = []  \ \land$\\
\hspace{10px} \ $(\forall opid.\ (\exists p.\ member \ (po \ p) \ opid) = 
    (opid \in (hd \ exe\_list)_{rops})) \ \land$\\
\hspace{10px} \ $(atomic\_flag\_val \ (hd \ exe\_list)_s = None) \ \land$\\
\hspace{10px} \ $(\forall i.\ i < (length \ exe\_list) - 1 \ \longrightarrow \ 
       po \ pbm \vdash_t (exe\_list!i) \leadsto (exe\_list!(i+1)))$
\end{tabular}
\end{definition}

We assume that $exe\_list$ must be non-empty because it must contain at least a witness $(hd \ exe\_list)$ for the initial setup. In the initial witness, the $xseq$ component must be an empty list, which means that nothing has been executed by the memory. The fourth conjunct states that the set of all operations is exactly the $rops$ component in the initial witness. The fifth conjunct ensures that the $atomic\_flag$ is $None$ initially. Lastly, each witness in $exe\_list$ is obtained by the transition $\vdash_t$. A \emph{final memory execution} is a valid execution witness list where every operation in the $rops$ part of the initial witness is in the $xseq$ part of the final witness. The keyword $set$ in the below definition converts a list to a set.
 
\begin{definition}[Valid Memory Execution Final Sequence]
\label{def:valid_mem_exe_final_seq}
\ \\
\begin{tabular}{l}
\hspace{10px}  $valid\_mem\_exe\_seq\_final \ po \ pbm \ exe\_list \equiv $\\
\hspace{10px} \ $valid\_mem\_exe\_seq \ po \ pbm \ exe\_list \ \land \ (hd \ exe\_list)_{rops} = set \ (last \ exe\_list)_{xseq}$
\end{tabular}
\end{definition}

The axiom Value is directly used in the operational model to obtain
load values, so the values we obtain for (atomic) load operations are
correct by definition. The soundness theorem is stated as follows
where we only check the other five axioms, and $(last \
exe\_list)_{xseq}$ is the $xseq$ part of the final witness:

\begin{flushleft}
\begin{tabular}{l}
\textbf{Theorem} $operational\_model\_sound:$ $valid\_mem\_exe\_seq\_final \ po \  pbm \ exe\_list \ \longrightarrow $\\ 
\ \ $(\forall opid \ opid'.\ axiom\_order \ opid \ opid' \ (last \ exe\_list)_{xseq} \ pbm) \ \land$\\
\ \ $(\forall opid.\ axiom\_termination \ opid \ po \ (last \ exe\_list)_{xseq} \ pbm) \ \land$\\
\ \ $(\forall opid \ opid'.\ axiom\_loadop \ opid \ opid' \ po \ (last \ exe\_list)_{xseq} \ pbm) \ \land$\\
\ \ $(\forall opid \ opid'.\ axiom\_storestore \ opid \ opid' \ po \ (last \ exe\_list)_{xseq} \ pbm) \ \land$\\
\ \ $(\forall opid \ opid'.\ axiom\_atomicity \ opid \ opid' \ po \ (last \ exe\_list)_{xseq} \ pbm)$
\end{tabular}
\end{flushleft}

\paragraph{Proof (outline)} (1) Axiom Order: assume that the two
memory operations $opid$ and $opid'$ are both store operations
($op\_type \ pbm \ opid/opid' \in \{st\_block, ast\_block\}$), they
are both in the $xseq$ part $(last \ exe\_list)_{xseq}$ of the final
witness, and $opid \neq opid'$. Then either $opid$ is before $opid'$
in the list $(last \ exe\_list)_{xseq}$ or $opid'$ is before $opid$.
By definition, the two operations must be related by $<_{(last
\ exe\_list)_{xseq}}$ in the memory order.

(2) Axiom Termination is satisfied by the construction of the
execution witness sequences, because the $xseq$ part of the final witness
is guaranteed to contain all the store operations, which means that
the execution of these operations have been completed by the memory.

(3) Axiom LoadOp: assume that $opid$ is a load ($ld\_block$ or
$ald\_block$), and $opid \ ;_{po}^p \ opid'$ where $p = proc\_of\_op
\ opid \ pbm$. We need to show that $opid \ <_{xseq} \ opid' = Some \
True$ where $xseq$ is the executed sequence of the final witness.
This is proved by contradiction. Assume the opposite that $opid \
<_{xseq} \ opid' = Some \ True$ does not hold. Since the sequence
$xseq$ in the final witness contains every memory operation, we
obtain that $opid' \ <_{xseq} \ opid = Some \ True$ must hold. Then
there must be a witness, e.g., the $i$th element $(xseq_i, rops_i,
s_i)$ in $exe\_list$, such that 
\begin{center}
$po \ pdm \ opid' \vdash_m xseq_i,s_i \leadsto xseq_{i+1},s_{i+1}$
\end{center}
holds, and $opid$ is not in $xseq_i$. By a case analysis, whatever the type of operation $opid'$ is, the last condition of the corresponding rule does not hold. Therefore we have a contradiction.

(4) Axiom StoreStore: proved similarly as above.

(5) Axiom Atomicity: suppose that $opid$ and $opid'$ are of type $ald
\_block$ and $ast\_block$ respectively, the $atom\_pair\_id$ field of
$opid'$ is $Some \ opid$, and $opid \ ;_{po}^p \ opid'$ holds for $p
= proc\_of\_op \ opid \ pbm$. We need to show that (5.1) $opid \
<_{xseq} \ opid' = Some \ True$ in the executed sequence $xseq$ of
the final witness, and (5.2) for every other store operation $opid''
\neq opid'$ such that $opid''$ is in $xseq$, either $opid'' \
<_{xseq} \ opid = Some \ True$ or $opid' \ <_{xseq} \ opid'' = Some \
True$. The argument for (5.1) is similar to the proof for (3). That
is, if the opposite is true, then condition 5 of the rule
$atom\_store$ would not hold when the operational model executes
$opid'$. For (5.2), assume the opposite that there is some $opid''$
in $xseq$ such that $opid \ <_{xseq} \ opid'' = Some \ True$ and
$opid'' \ <_{xseq} \ opid' = Some \ True$. Then we can find three
witnesses, which are the $i$th, $j$th, and $k$th element in
$exe\_list$ respectively, such that 
\begin{center}
$po \ pbm \ opid \vdash_m xseq_i,s_i \leadsto xseq_{i+1},s_{i+1}$\\
$po \ pbm \ opid'' \vdash_m xseq_j,s_j \leadsto xseq_{j+1},s_{j+1}$\\
$po \ pbm \ opid' \vdash_m xseq_k,s_k \leadsto xseq_{k+1},s_{k+1}$
\end{center}
hold, and $i < j$ and $j < k$. By the types of operations, the first operation for $opid$ is obtained by the rule $atom\_load$, which sets the $atomic\_flag$ to $Some \ opid$ in the end. An inspection of the rules shows that only the rule $atom\_store$ that applies on the $opid'$ operation which pairs with $opid$ can set the $atomic\_flag$ back to $None$. Since each instance of $op\_id$ is unique, we know that there is only one store operation that pairs with $opid$. Therefore any witness between $i$ and $k$ will see $atomic\_flag$ as $Some \ opid$, including the $j$th one. Since we assume that $opid''$ is a store, it can only be $st\_block$ or $ast\_block$, and the rule that obtains the second operation can only be $store$ or $atom\_store$. If it is the former, condition 3 in the rule $store$ is violated. Otherwise, condition 3 and 4 of the rule $atom\_store$ are violated, because the corresponding atomic load part of $opid''$ cannot be $opid$.\qed 
 
\subsection{Completeness of the Operational TSO Model}
\label{subsec:op_comp}

This subsection shows that the operational TSO model can produce a
final memory execution witness list for every final sequence of
memory operations that satisfies the TSO axioms.

In contrast to the definition of valid memory execution sequences
(cf. Definition~\ref{def:valid_mem_exe_seq}
and~\ref{def:valid_mem_exe_final_seq}), which describe memory
operation sequences constructed by the TSO operational memory model,
we now define a final sequence $op\_seq$ (of type $op\_id \ list$) of
arbitrary memory operations that satisfies certain conditions. This
section shows that every such $op\_seq$ that satisfies the TSO axioms
can be constructed by the operational TSO model.

\begin{definition}[Valid Memory Operation Final Sequence]
\label{def:valid_mem_op_final_seq}
\ \\
\begin{tabular}{l}
\hspace{10px} $op\_seq\_final \ \ op\_seq \ po \ pbm \equiv valid\_program \ po \ pbm \ \land $\\
\hspace{10px} \ $length \ op\_seq \ > \ 0 \ \land \ (\forall opid.\ (\exists p.\ member \ (po \ p) \ opid) = (member \ op\_seq \ opid)) \ \land$\\
\hspace{10px} \ $non\_repeat\_list \ op\_seq \ \land \ (\forall opid.\ (\exists p.\ member \ (po \ p) \ opid) \ \longrightarrow $\\ 
\hspace{10px} \ \ \ $(op\_type \ (pbm \ opid) \in \{ld\_block, st\_block, ald\_block, ast\_block\}))$
\end{tabular}
\end{definition}

\begin{figure}[t!]
	\centering
	\begin{tabular}{c}
		\AxiomC{$type_{id} = ld$}
		\AxiomC{$\forall id'. \ ((id' \ ; \ id) \ \land \ 
			type_{id'} \in \{ld,ald\} \longrightarrow \ id'\in x)$}
		\alwaysSingleLine
		\RightLabel{\scriptsize $load$}
		\BinaryInfC{$x,s \leadsto x@[id],(exe^{last}_{id} \ Lval_{id}  \ (exe^{pre}_{id}  \ s))$}
		\DisplayProof\\[30px]
		\AxiomC{$type_{id} = st$}
		\alwaysNoLine
		\AxiomC{$flag_{atom} = undefined$}
		\BinaryInfC{$\forall id'. ((id' \ ; \ id) \ \land \ type_{id'} \in \{ld, ald, st, ast\} \longrightarrow id'\in x)$}
		\alwaysSingleLine
		\RightLabel{\scriptsize $store$}
		\UnaryInfC{$x,s \leadsto x@[id],(W_{mem} \ id \ (exe_{id} \ s))$}
		\DisplayProof\\[30px]
		\AxiomC{$type_{id} = ald$}
		\alwaysNoLine
		\AxiomC{$flag_{atom} = undefined$}
		\BinaryInfC{$\forall id'. ((id' \ ; \ id) \ \land \ type_{id'} \in \{ld, ald, st, ast\} \longrightarrow id'\in x)$}
		\RightLabel{\scriptsize $atom\_load$}
		\alwaysSingleLine
		\UnaryInfC{$x,s \leadsto x@[id],(flag^{set}_{atom} \ id \ (exe^{last}_{id} \ Lval_{id}  \ (exe^{pre}_{id}  \ s)))$} 
		\DisplayProof\\[30px]
		\AxiomC{$type_{id} = ast$}
		\alwaysNoLine
		\AxiomC{$flag_{atom} = id'$}
		\AxiomC{$atom_{pair} \ id = id'$}
		\TrinaryInfC{$\forall id''. ((id'' \ ; \ id) \ \land \ type_{id''} \in \{ld, ald, st, ast\} \longrightarrow id''\in x)$} 
		\RightLabel{\scriptsize $atom\_store$}
		\alwaysSingleLine
		\UnaryInfC{$x,s \leadsto x@[id],(W_{mem} \ id \ (flag^{set}_{atom} \ undef \ (exe_{id} \ s)))$}
		\DisplayProof
	\end{tabular}
	\caption{Rules for the operational TSO model.}
	\label{fig:op_tso_model}
\end{figure}

\noindent We assume that $po \ pbm$ yield a valid program, and $op\_seq$ is non-empty, because we are not interested in nil executions. The third conjunct means that $op\_seq$ includes all the memory operations issued by all processors. The fourth conjunct reinforces that $op\_id$s in $op\_seq$ are unique. The last condition restricts our attention to memory operations. That is, we do not consider blocks of type $o\_block$ here, because their order is not constrained by the memory model. In an execution, the operational model guarantees that $o\_block$s are executed the last by their corresponding processors. The order between two $o\_block$ operations from different processors is not our concern because it has no effect on the result of execution.

The outline of the completeness proof is an induction on the length
of $op\_seq$. The core part is as follows: suppose we have a partial
execution witness list $exe\_list$ which corresponds to a partial
sequence of memory operations $sub\_op\_seq$ such that $\exists l.\
sub\_op\_seq@l = op\_seq$ holds. We show that for any memory
operation $opid$, if no rules in Figure~\ref{fig:op_tso_model} are
applicable to execute $opid$ as the next operation of $exe\_list$,
then no matter what the remaining operation sequence $l'$ is,
$sub\_op\_seq@[opid]@l'$ must violate at least one of the following
axioms: LoadOp, StoreStore, and Atomicity. This part of the proof is
divided into four sub-proofs depending on the type of $opid$. For
space reasons we only discuss the main lemmas here.

The first lemma shows that if a load operation $opid$ cannot be
executed as the next step, then the axiom LoadOp does not hold
for the executed operation list $(last \ exe\_list)_{xseq}@[opid]$.

\begin{flushleft}
\begin{tabular}{l}
\textbf{Lemma} $load\_op\_contra\_violate: \ valid\_mem\_exe\_seq \ po \ pbm \ exe\_list \ \longrightarrow$\\
\ \ $op\_type \ (pbm \ opid) = ld\_block \ \longrightarrow$\\
\ \ $\not\exists xseq' \ s'.\ (po \ pbm \ opid \vdash_m \ (last \ exe\_list)_{xseq},(last \ exe\_list)_s \leadsto xseq',s') \ \longrightarrow$\\
\ \ $\exists opid'.\ (axiom\_loadop\_violate \ opid' \ opid \ po \ ((last \  exe\_list)_{xseq}@[opid]) \ pbm)$
\end{tabular}
\end{flushleft}

\paragraph{Proof (outline)} If $opid$ is a $ld\_block$, and the rule
$load$ is not applicable on $exe\_list$ and $opid$, by the conditions
of the rule, there must exist some $opid'$ such that $opid' \
;_{po} \ opid$, $op\_type \ opid' \in
\{ld\_block, ald\_block\}$, and $\lnot (member \ xseq \ opid')$,
where $xseq = (last \ exe\_list)_{xseq}$. Then we obtain
that $(opid' \ <_{xseq@[opid]} \ opid) = Some \ False$. Therefore $opid'$
and $opid$ violates the axiom LoadOp on the sub-sequence
$xseq@[opid]$. \qed 

If a store operation $opid$ cannot be executed in the next step, then
at least one of LoadOp, StoreStore, and Atomicity does not hold for
the sequence $(last \ exe\_list)_{xseq}@[opid]$.

\begin{flushleft}
\begin{tabular}{l}
\textbf{Lemma} $store\_op\_contra\_violate: \ valid\_mem\_exe\_seq \ po \ pbm \ exe\_list \ \longrightarrow$\\
\ \ $op\_type \ (pbm \ opid) = st\_block \ \longrightarrow $\\
\ \ $\not\exists xseq' \ s'.\ (po \ pbm \ opid \vdash_m \ (last \ exe\_list)_{xseq},(last \ exe\_list)_s \leadsto xseq',s') \ \longrightarrow$\\
\ \ $\lnot (member \ (last \ exe\_list)_{xseq} \ opid) \ \longrightarrow$\\
\ \ $op\_seq\_final \ (((last \ exe\_list)_{xseq}@[opid])@remainder) \ po \ pbm \ \longrightarrow$\\
\ \ $\exists opid' \ opid''.\ (axiom\_loadop\_violate \ opid' \ opid'' \ po \ ((last \ exe\_list)_{xseq}@[opid]) \ pbm) \ \lor$\\
\ \ \ \ $(axiom\_storestore\_violate \ opid' \ opid'' \ po \ ((last \ exe\_list)_{xseq}@[opid]) \ pbm) \ \lor$\\
\ \ \ \ $(axiom\_atomicity\_violate \ opid' \ opid'' \ po \ ((last \ exe\_list)_{xseq}@[opid]) \ pbm)$
\end{tabular}
\end{flushleft}

\paragraph{Proof (outline)} If $opid$ is a $st\_block$ and the rule
$store$ is not applicable, by the conditions of the rule, either (1)
$atomic\_flag\_val \ (last \ exe\_list)_s \neq None$,
or (2) there exists some $opid'$ such that $opid' \
;_{po} \ opid$, $op\_type \ opid' \in
\{ld\_block, ald\_block, st\_block, ast\_block\}$, and $\lnot (member
\ xseq \ opid')$, where $xseq = (last \ exe\_list)_{xseq}$.

For the first case, there must be some atomic load operation $opid''$
such that the $atomic\_flag$ in this state is $Some \ opid''$. This
implies that the atomic load part of an atomic load-store instruction
has been executed by the memory, but the atomic store part has not
been executed. Then the sub-sequence $xseq@[opid]$ already violates
the axiom Atomicity, because the store operation $opid$ is after the
atomic load $opid''$, but it is before the corresponding atomic store
part.

For the second case, we obtain that $opid' \ <_{xseq@[opid]} \ opid =
Some \ False$. If $opid'$ is a $ld\_block$ or a $ald\_block$, then
the axiom LoadOp is violated by $opid'$, $opid$, and the sub-sequence
$xseq@[opid]$. If $opid'$ is a $st\_block$ or a $ast\_block$, then
the axiom StoreStore is violated. \qed 

The lemma for atomic load operations requires reasoning about all
possible future execution sequences. We show that if an atomic load
operation $opid$ cannot be executed as the next step, then either
LoadOp does not hold for $(last \ exe\_list)_{xseq}@[opid]$, or at
least one of StoreStore and Atomicity does not hold for $(last
\ exe\_list)_{xseq}@[opid]@remainder$, for any extension $reminder$ of the
next step that forms a final $op\_seq$.

\begin{flushleft}
\begin{tabular}{l}
\textbf{Lemma} $atom\_load\_op\_contra\_violate: \ valid\_mem\_exe\_seq \ po \ pbm \ exe\_list \ \longrightarrow$\\
\ \ $op\_type \ (pbm \ opid) = ald\_block \ \longrightarrow $\\
\ \ $\not\exists xseq' \ s'.\ (po \ pbm \ opid \vdash_m \ (last \ exe\_list)_{xseq},(last \ exe\_list)_s \leadsto xseq',s') \ \longrightarrow$\\
\ \ $member \ (po \ (proc\_of\_op \ opid \ pbm)) \ opid \ \longrightarrow$ \\
\ \ $\lnot (member \ (last \ exe\_list)_{xseq} \ opid) \ \longrightarrow$\\
\ \ $op\_seq\_final \ (((last \ exe\_list)_{xseq}@[opid])@remainder) \ po \ pbm \ \longrightarrow$\\
\ \ $\exists opid' \ opid''.\ (axiom\_loadop\_violate \ opid' \ opid'' \ po \ ((last \ exe\_list)_{xseq}@[opid]) \ pbm) \ \lor$\\
\ \ \ \ $(axiom\_storestore\_violate \ opid' \ opid'' \ po \ ((last \ exe\_list)_{xseq}@[opid]@remainder) \ pbm) \ \lor$\\
\ \ \ \ $(axiom\_atomicity\_violate \ opid' \ opid'' \ po \ ((last \ exe\_list)_{xseq}@[opid]@remainder) \ pbm)$
\end{tabular}
\end{flushleft}

\paragraph{Proof (outline)} If $opid$ is a $ald\_block$ and the rule
$atom\_load$ is not applicable, by the conditions of the rule, either (1)
$atomic\_flag\_val \ (last \ exe\_list)_s \neq None$,
or (2) there exists some $opid'$ such that $opid' \
;_{po} \ opid$, $op\_type \ opid' \in
\{ld\_block, ald\_block, st\_block, ast\_block\}$, and $\lnot (member
\ xseq \ opid')$, where $xseq = (last \ exe\_list)_{xseq}$.

For the first case, there must be some $opid''$ such that the
$atomic\_flag$ in this state is $Some \ opid''$. So $opid''$ is a
$ald\_block$ that has been executed by the memory, and its corresponding
atomic store part $opid''_{ast}$ has not been executed. Suppose we
execute $opid$ now. Since $opid$ is also an atomic load block, it
must have a corresponding atomic store block $opid_{ast}$, which has
not been executed. In the remaining execution, if $opid''_{ast}$ is
executed before $opid_{ast}$, then $opid''_{ast}$ is in between
$opid$ and $opid_{ast}$, which violates the axiom Atomicity. Otherwise
$opid_{ast}$ is executed before $opid''_{ast}$, then $opid_{ast}$ is
in between $opid''$ and $opid''_{ast}$, which again violates the axiom
Atomicity. Therefore it does not matter what $remainder$ is,
$xseq@[opid]@remainder$ always falsifies Atomicity.

For the second case, we obtain that $opid' \ <_{xseq@[opid]} \ opid =
Some \ False$. If $opid'$ is a $ld\_block$ or a $ald\_block$, then
the axiom LoadOp is falsified. If $opid'$ is a $st\_block$ or
$ast\_block$, we consider two sub-cases: (2.1) if $opid'$ is executed
before the atomic store block $opid_{ast}$ corresponding to $opid$,
then Atomicity is violated. (2.2) Otherwise $opid'$ is executed after
$opid_{ast}$. Since $opid' \ ;_{po} opid$ and, if the program is
valid, $opid \ ;_{po} \ opid_{ast}$, by transitivity of $;$, we
obtain that $opid' \ ;_{po} \ opid_{ast}$. Then we have that
StoreStore is violated by $opid'$, $opid_{ast}$, and
$xseq@[opid]@remainder$ for any $remainder$ such that
$xseq@[opid]@remainder = op\_seq$. \qed 

The lemma for atomic store operations also involves reasoning about
all possible future executions, but only for the axiom Atomicity.

\begin{flushleft}
\begin{tabular}{l}
\textbf{Lemma} $atom\_store\_op\_contra\_violate: \ valid\_mem\_exe\_seq \ po \ pbm \ exe\_list \ \longrightarrow$\\
\ \ $op\_type \ (pbm \ opid) = ast\_block \ \longrightarrow $\\
\ \ $\not\exists xseq' \ s'.\ (po \ pbm \ opid \vdash_m \ (last \ exe\_list)_{xseq},(last \ exe\_list)_s \leadsto xseq',s') \ \longrightarrow$\\
\ \ $member \ (po \ (proc\_of\_op \ opid \ pbm)) \ opid \ \longrightarrow$ \\
\ \ $\lnot (member \ (last \ exe\_list)_{xseq} \ opid) \ \longrightarrow$\\
\ \ $op\_seq\_final \ (((last \ exe\_list)_{xseq}@[opid])@remainder) \ po \ pbm \ \longrightarrow$\\
\ \ $\exists opid' \ opid''.\ (axiom\_loadop\_violate \ opid' \ opid \ po \ ((last \ exe\_list)_{xseq}@[opid]) \ pbm) \ \lor$\\
\ \ \ \ $(axiom\_storestore\_violate \ opid' \ opid \ po \ ((last \ exe\_list)_{xseq}@[opid]) \ pbm) \ \lor$\\
\ \ \ \ $(axiom\_atomicity\_violate \ opid' \ opid'' \ po \ ((last \ exe\_list)_{xseq}@[opid]@remainder) \ pbm)$
\end{tabular}
\end{flushleft}

\paragraph{Proof (outline)} If $opid$ is a $ast\_block$ and the rule
$atom\_store$ is not applicable, by the conditions of the rule,
either (1) $atomic\_flag\_val \ (last \ exe\_list)_s
= None$, (2) $atomic\_flag\_val$  $(last \
exe\_list)_s \neq atom\_pair\_id \ (pbm \ opid)$, or (3) there exists
some $opid'$ such that $opid' \ ;_{po} \ opid$, $op\_type \ opid' \in
\{ld\_block, ald\_block, st\_block, ast\_block\}$, and $\lnot (member
\ xseq \ opid')$, where $xseq$ $=$ $(last \ exe\_list)_{xseq}$.

If the program is valid, $opid$ must have a corresponding atomic load
block $opid_{ald}$ and $atom\_pair\_id$ $(pbm \ opid) = Some \
opid_{ald}$. Thus the first case is a sub-case of the second one. For
the second case, if $opid_{ald}$ has not been executed by the memory,
then $xseq@[opid]$ violates Atomicity because the atomic store part
is executed before the atomic load part. Otherwise, $opid_{ald}$ has
been executed by the memory. The rule $atom\_load$ ensures that after
executing $opid_{ald}$, the $atomic\_flag$ of the state is set as
$Some \ opid_{ald}$. The other rules ensure that the $atomic\_flag$
remains unchanged until the memory execution of $opid$. Therefore in
the current state $atomic\_flag$ must still be $Some \ opid_{ald}$,
contradicting the second case. The third case is analogous to the
second case of the proof for Lemma~$store\_op\_contra\_violate$. \qed 

Some parts of the above proofs reason about future executions, while
others only show that a sub-sequence violates an axiom. We strengthen
this result by showing that the violation ``persists'': once a
sequence of memory operations violates an axiom, any extension of it
also violates the axiom. This is proved by the three lemmas below.

\begin{flushleft}
\begin{tabular}{l}
\textbf{Lemma} $axiom\_loadop\_violate\_persist: \ valid\_program \ po \ pbm \ \longrightarrow$\\
\ \ $axiom\_loadop\_violate \ opid \ opid' \ po \ xseq \ pbm \longrightarrow \ 
non\_repeat\_list \ (xseq@xseq')$\\
\ \ $axiom\_loadop\_violate \ opid \ opid' \ po \ (xseq@xseq') \ pbm$
\end{tabular}
\end{flushleft}

\begin{flushleft}
\begin{tabular}{l}
\textbf{Lemma} $axiom\_storestore\_violate\_persist: \ valid\_program \ po \ pbm \ \longrightarrow$\\
\ \ $axiom\_storestore\_violate \ opid \ opid' \ po \ xseq \ pbm \longrightarrow \ 
non\_repeat\_list \ (xseq@xseq')$\\
\ \ $axiom\_storestore\_violate \ opid \ opid' \ po \ (xseq@xseq') \ pbm$
\end{tabular}
\end{flushleft}

\begin{flushleft}
\begin{tabular}{l}
\textbf{Lemma} $axiom\_atomicity\_violate\_persist: \ valid\_program \ po \ pbm \ \longrightarrow$\\
\ \ $axiom\_atomicity\_violate \ opid \ opid' \ po \ xseq \ pbm \longrightarrow \ 
non\_repeat\_list \ (xseq@xseq')$\\
\ \ $axiom\_atomicity\_violate \ opid \ opid' \ po \ (xseq@xseq') \ pbm$
\end{tabular}
\end{flushleft}

Combining the above proofs, we obtain that if a memory operation
$opid$ cannot be executed by the operational model as the next step,
then any future execution that includes $opid$ as the next step must
falsify the axiomatic TSO model. This is formalised in the lemma
below:

\begin{flushleft}
\begin{tabular}{l}
\textbf{Lemma} $mem\_op\_contra\_violate: \ valid\_mem\_exe\_seq \ po \ pbm \ exe\_list \ \longrightarrow$\\
\ \ $op\_seq\_final \ (((last \ exe\_list)_{xseq}@[opid])@remainder) \ po \ pbm \ \longrightarrow$\\
\ \ $\not\exists xseq' \ s'.\ (po \ pbm \ opid \vdash_m \ (last \ exe\_list)_{xseq},(last \ exe\_list)_s \leadsto xseq',s') \ \longrightarrow$\\
\ \ $\exists opid' \ opid''.$\\ 
\ \ \ \ $(axiom\_loadop\_violate \ opid' \ opid'' \ po \ ((last \ exe\_list)_{xseq}@[opid]@remainder) \ pbm) \ \lor$\\
\ \ \ \ $(axiom\_storestore\_violate \ opid' \ opid'' \ po \ ((last \ exe\_list)_{xseq}@[opid]@remainder) \ pbm) \ \lor$\\
\ \ \ \ $(axiom\_atomicity\_violate \ opid' \ opid'' \ po \ ((last \ exe\_list)_{xseq}@[opid]@remainder) \ pbm)$
\end{tabular}
\end{flushleft}

Taking the contrapositive of the above, we obtain that for any $opid$ and $remainder$ such that $sub\_op\_seq@[opid]@remainder = op\_seq$, if $op\_seq$
satisfies the TSO axioms, then there is a rule application that
executes $opid$ as the next step from the last state in $exe\_list$. This concludes the core lemma of the completeness theorem:

\begin{flushleft}
\begin{tabular}{l}
\textbf{Lemma} $mem\_op\_exists: \ valid\_mem\_exe\_seq \ po \ pbm \ exe\_list \ \longrightarrow$\\
\ \ $op\_seq\_final \ (((last \ exe\_list)_{xseq}@[opid])@remainder) \ po \ pbm \ \longrightarrow$\\
\ \ $(\forall opid' \ opid''.$\\ 
\ \ \ \ $(axiom\_loadop \ opid' \ opid'' \ po \ ((last \ exe\_list)_{xseq}@[opid]@remainder) \ pbm) \ \land$\\
\ \ \ \ $(axiom\_storestore \ opid' \ opid'' \ po \ ((last \ exe\_list)_{xseq}@[opid]@remainder) \ pbm) \ \land$\\
\ \ \ \ $(axiom\_atomicity \ opid' \ opid'' \ po \ ((last \ exe\_list)_{xseq}@[opid]@remainder) \ pbm)) \ \longrightarrow$\\
\ \ $\exists xseq' \ s'.\ (po \ pbm \ opid \vdash_m \ (last \ exe\_list)_{xseq},(last \ exe\_list)_s \leadsto xseq',s')$
\end{tabular}
\end{flushleft}

Finally, with a careful setup of the initial witness, the
completeness theorem can be proven by an induction in which the above
lemma is used to obtain the execution witnesses at each step.

\begin{flushleft}
\begin{tabular}{l}
\textbf{Theorem} $operational\_model\_complete:$ 
  $op\_seq\_final \ op\_seq \ po \ pbm \longrightarrow $\\
\ \ $(\forall opid \ opid'.\ axiom\_loadop \ opid \ opid' \ po \ op\_seq \ pbm) \ \longrightarrow $\\ 
\ \ $(\forall opid \ opid'.\ axiom\_storestore \ opid \ opid' \ po \ op\_seq \ pbm) \ \longrightarrow $\\
\ \ $(\forall opid \ opid'.\ axiom\_atomicity \ opid \ opid' \ po \ op\_seq \ pbm) \ \longrightarrow$\\ 
\ \ $(\exists exe\_list.\ valid\_mem\_exe\_seq\_final \ po \ pbm \ exe\_list \ \land \ (last \ exe\_list)_{xseq} = op\_seq)$
\end{tabular}
\end{flushleft}